\documentclass[sigconf, nonacm]{acmart}
\usepackage{listings}
\usepackage{subfig}

\AtBeginDocument{
  \providecommand\BibTeX{{
    \normalfont B\kern-0.5em{\scshape i\kern-0.25em b}\kern-0.8em\TeX}}}

\begin{document}

\title{OmniBuds: A Sensory Earable Platform for Advanced Bio-Sensing and On-Device Machine Learning}

\author{\large Alessandro Montanari \hspace{0.25em} Ashok Thangarajan \hspace{0.25em} Khaldoon Al-Naimi \hspace{0.25em} Andrea Ferlini \hspace{0.25em} Yang Liu \hspace{0.25em} Ananta Narayanan Balaji \hspace{0.25em} Fahim Kawsar}
\affiliation{
 \institution{\small Nokia Bell Labs, Cambridge (UK) }
 \country{}
}

\renewcommand{\shortauthors}{Montanari et al.}

\settopmatter{printacmref=false}

\begin{abstract}
Sensory earables have evolved from basic audio enhancement devices into sophisticated platforms for clinical-grade health monitoring and wellbeing management. This paper introduces OmniBuds, an advanced sensory earable platform integrating multiple biosensors and onboard computation powered by a machine learning accelerator, all within a real-time operating system (RTOS). The platform's dual-ear symmetric design, equipped with precisely positioned kinetic, acoustic, optical, and thermal sensors, enables highly accurate and real-time physiological assessments.
Unlike conventional earables that rely on external data processing, OmniBuds leverage real-time onboard computation to significantly enhance system efficiency, reduce latency, and safeguard privacy by processing data locally. This capability includes executing complex machine learning models directly on the device. We provide a comprehensive analysis of OmniBuds' design, hardware and software architecture demonstrating its capacity for multi-functional applications, accurate and robust tracking of physiological parameters, and advanced human-computer interaction.

\end{abstract}


\maketitle

\section{Introduction}
Sensory earables have transcended their initial promise of enhanced audio experiences to become powerful tools for accurate health monitoring and personal wellbeing management. Central to this transformation is their ability to leverage the ear’s unique anatomical proximity to critical vascular and acoustic structures, enabling precise physiological sensing while minimising the effects of motion artefacts~\cite{roddiger2022sensing, choudhury2021earable}.

The integration of advanced biosensors capable of tracking accurate vital markers allows earables to offer a comprehensive assessment of an individual’s physical, physiological, and social context. Recent academic studies have revealed the extensive potential of these devices in a variety of applications. Earables have demonstrated significant promise in the precise tracking of fitness metrics~\cite{prakash2019stear, ferlini2019head, ferlini2021eargate, atallah2014gait, ferlini2021enabling}, monitoring cardiovascular and respiratory parameters~\cite{ferlini2021ear, romero2024optibreathe, butkow2023heart}, hearing screening~\cite{shahid2024towards, demirel2023cancelling, chan2023wireless} and modelling motor symptoms associated with neurological disorders such as Parkinson's disease~\cite{goverdovsky2017hearables, bleichner2017concealed, kidmose2013study, kalanadhabhatta2021fatigueset}. They have also been utilised to support dementia patients through cognitive assistance~\cite{franklin2021designing} and to enhance auditory function via augmented hearing~\cite{veluri2024ai, veluri2024look, veluri2023semantic, yang2020ear, yang2021personalizing, demirel2024unobtrusive}. Additionally, earables are pioneering the next frontier of personal computing devices, advancing human-computer interaction (HCI) by enabling seamless and intuitive engagement with devices worn on or around the body~\cite{roddiger2022sensing}. This growing body of research underscores the transformative role of earables, particularly in healthcare and HCI, as they pave the way for innovative developments in personal computing.

Yet, despite remarkable strides, the ultimate challenge remains: creating an earable platform that seamlessly combines high performance sensing, functional utility, and design efficiency within the compact form factor demanded by the ear. It is within this delicate balance that the next breakthrough in sensory earables lies.

In 2018, we introduced eSense~\cite{kawsar2018earables}, a multisensory in-ear platform aimed at advancing research into intelligent earables. By integrating motion and audio sensors with Bluetooth Low Energy (BLE) connectivity, eSense facilitated the monitoring of activities such as speech and movement, while offering APIs to enable developers to access sensor data. This platform provided a foundation for exploring innovative capabilities; however, it also highlighted key challenges, including the need for enhanced sensor accuracy, system integration, design optimisation, user experience, and societal acceptance. These findings revealed both the limitations of eSense and its significant potential for further development and wider application in earable technology.

\begin{figure}
    \centering
    \includegraphics[width=1.0\columnwidth]{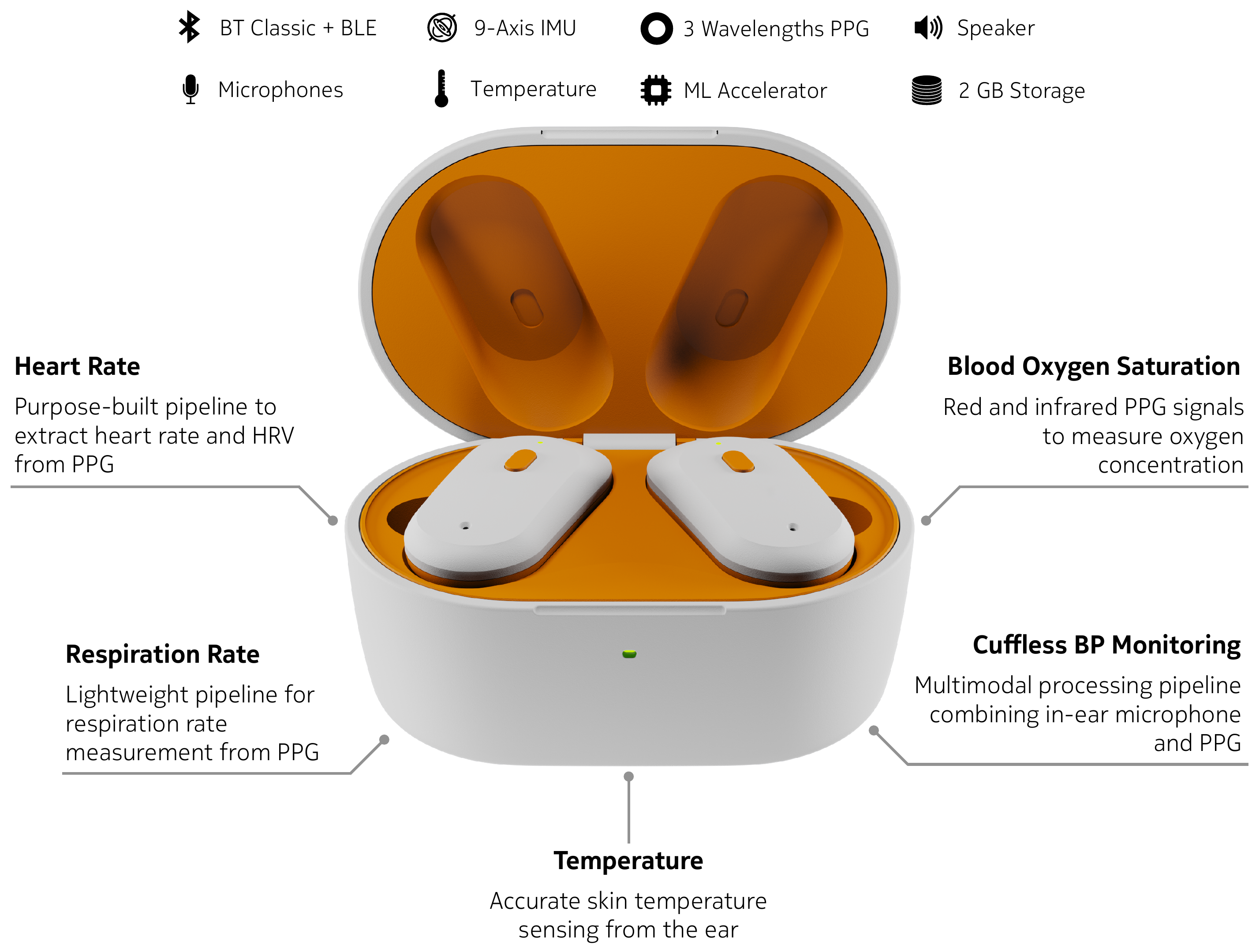}
    \vspace{-8mm}
    \caption{OmniBuds main HW components and vital signs monitored.}
    \label{fig:OmnibudImage}
    \vspace{-5mm}
\end{figure}

Our reflection suggests that the true utility of sensory earables hinges on their ability to deliver precise and reliable sensing, a task that demands careful sensor placement to fully exploit the anatomical advantages of the ear. The inherent symmetry of the ears offers a unique opportunity, where spatial redundancy can significantly enhance the accuracy of physiological observations. Expanding beyond kinetic and acoustic sensors to incorporate advanced modalities such as optical PPG and temperature sensors introduces an array of new possibilities for health and wellbeing monitoring. In dual-ear configurations, this spatial redundancy unlocks multi-dimensional, robust health assessments, vastly increasing the scope of applications for these devices.

Currently, platforms like eSense operate as passive data collectors, with no onboard computation, relying entirely on external systems for processing. This design introduces inefficiencies, including communication bottlenecks, increased latency, and critical privacy risks due to the need for constant data transmission. We envision integrating local computational capabilities and sufficient storage within ergonomically designed earables. This shift would enable real-time processing, facilitate the execution of machine learning models directly on the device, and drastically enhance system efficiency and responsiveness. Privacy concerns would also be mitigated, as sensitive data could remain on the device without being transmitted externally. Moreover, the inclusion of powerful and programmable digital signal processing (DSP) can further amplify their impact, enabling augmented auditory experiences, spatial hearing enhancements, and even therapeutic interventions.

We anticipate that, with these advancements, earables can evolve from simple firmware-driven devices into fully developed programmable software platforms. This transformation would allow multiple applications to coexist, efficiently multitask, and optimally access sensors, creating a rich, integrated user experience untethered from secondary devices. However, advanced sensing, computing, and intervention capabilities must not compromise the primary user experience—the core functionality of earables as wireless audio devices. Therefore, a unified and efficient communication framework is essential to seamlessly support both wireless audio and these additional functionalities, ensuring a cohesive and high-quality user experience without interference.

These advancements represent a significant shift in the development of earables, positioning them at the forefront of personal health technology and human-computer interaction. In this paper, we introduce OmniBuds, a state-of-the-art sensory earable platform that exemplifies this shift. With an advanced array of carefully placed sensors, onboard computation powered by a machine learning accelerator, local storage, and programmable digital signal processors (DSPs), OmniBuds is designed to operate within a software stack built on a real-time operating system (RTOS). Its functionally and spatially symmetric dual-ear design opens new opportunities in accurate bio-sensing, targeted interventions, and multi-application functionality, all while ensuring ultra-efficient and privacy-preserving operation in a compact and ergonomic form factor.

The paper first outlines the design principles behind OmniBuds, followed by a detailed explanation of its hardware and software architecture. Each key subsystem is examined demonstrating the platform's capabilities. Finally, we present applications of OmniBuds and explore their vast potential for future innovation, emphasising their transformative role in the advancement of earable technology.

\section{Design Principles}
\label{sec:principles}

The design of OmniBuds was driven by a central philosophy: integrating hardware and software in a way that maximises performance, efficiency, and versatility for earable computing. Unlike many other devices, where hardware and software development may occur in silos, OmniBuds was conceived as a cohesive platform where the two elements are deeply intertwined. 

At the core of this design is the recognition that hardware choices dictate software capabilities, and vice versa. This interdependence enabled us to develop a platform that meets the current technical requirements of researchers. Additionally, it is designed to anticipate future needs in the field, particularly in terms of scalability, flexibility, and real-time processing.

In the following sections, we will explore the key design principles that guided the design of OmniBuds, highlighting how these principles shaped both the device's hardware and software.

\subsection{Local Computation and Energy Efficiency}
A fundamental principle guiding the design of OmniBuds is the prioritisation of local computation. Unlike conventional earbuds, which often rely on external devices to handle intensive processing tasks, OmniBuds are designed to manage complex workloads directly on the device. This approach significantly reduces the need for constant communication with external systems, a major source of power consumption in wearable devices.

For instance, integrating a CNN accelerator enables OmniBuds to run sophisticated machine learning models on-device, such as those used for speech recognition. 
The power of the CNN accelerator lies in its ability to execute these complex models in milliseconds, without needing to stream audio data to an external device for processing. This enhances the responsiveness of the earbuds while also significantly reducing battery drain by keeping all computations local. The ability to process speech directly within the device enables applications where real-time interaction and low-latency responses are critical, such as hands-free control of devices or real-time assistance in challenging environments\cite{veluri2024ai, veluri2024look, veluri2023semantic}.

Following this principle enables OmniBuds to support a wide range of applications, from real-time biometric monitoring to interactive audio experiences, all while ensuring extended battery life and data privacy.

\subsection{Dynamic Sensor Access and Multitasking}
One of the central design principles of OmniBuds is the integration of sensor flexibility and multitasking~\cite{min2022sensix, min2023sensix++}, which is achieved by tightly coupling hardware and software design. The platform leverages intelligent sensors capable of performing many functions autonomously, reducing the need to constantly wake up the main microcontroller (MCU). These sensors are equipped with built-in hardware functionalities, such as basic data processing and event detection, allowing them to handle critical tasks independently. 

On the software side, OmniBuds supports a multitasking environment that allows multiple applications to access sensor data simultaneously. Rather than restricting sensor access to a single application at a time, the system is designed to allow multiple applications to share sensor data without conflicts or excessive resource consumption. This requires a coordinated approach at the system level, where the OmniBuds software system manages access and ensures that each application can retrieve the data it needs without impacting the performance of other applications.
For instance, physiological monitoring applications can access PPG sensor data while another application may simultaneously use the same data stream for user authentication~\cite{yadav2018evaluation} or facial expression detection~\cite{choi2022ppgface}. The system efficiently manages this data access, ensuring that sensor data is shared across applications without compromising performance or power efficiency.

This holistic design—integrating flexible, autonomous sensors with multitasking software—ensures that OmniBuds remain adaptable to a wide range of experimental or real-world applications. Researchers and developers can take full advantage of the platform's ability to run multiple, independent applications simultaneously, maximising the potential of each sensor and minimising unnecessary power consumption.

\subsection{Unified Communication Across Devices}
A key design principle of OmniBuds is to enable seamless communication between devices and software modules, ensuring that the platform remains both scalable and developer-friendly. This capability is underpinned by the integration of a Bluetooth Classic and Bluetooth Low Energy (BLE) chip at the hardware level and a unified set of APIs at the software level. This integration allows OmniBuds to abstract away the complexities of device-to-device communication, enabling modules to interact seamlessly, whether running locally on the earbuds or across multiple devices in the ecosystem.

The unified communication framework simplifies multi-device coordination while providing developers with a consistent interface for building robust and scalable applications. This framework allows applications to communicate across earbuds, or with external systems, without requiring developers to manage the intricacies of local versus remote execution. For example, real-time data from one earbud can be transmitted to the other or to an external device, such as a smartphone, while maintaining a high level of performance and low latency.

OmniBuds’ communication capabilities are particularly suited for research scenarios where multiple devices may need to work in concert. Whether managing sensor data, synchronising tasks, or coordinating between devices, the unified BLE communication framework ensures that developers can focus on building applications rather than dealing with the complexities of communication protocols.

\subsection{Privacy by Design}
In modern wearable systems, especially those handling sensitive physiological data, privacy is a non-negotiable requirement. OmniBuds address this through a dual-layer approach, combining local data processing at the hardware level with secure communication channels in the software.

By prioritising local computation, OmniBuds minimise the need to transmit sensitive data externally, reducing the risk of unauthorised access. When external communication is necessary, the platform uses encrypted Bluetooth Low Energy (BLE) to protect data integrity. This approach ensures that OmniBuds meet the stringent privacy requirements of health-focused research, safeguarding user data throughout all operations.

Demonstrating the synergy between hardware and software, OmniBuds set a new standard in earable technology. Designed to address current research demands and anticipate future needs, they combine seamless hardware-software integration with a strong focus on privacy and flexibility, positioning them as a leader in earable computing research.

\section{Hardware Architecture}
\label{sec:hw_arch}

\begin{figure}
    \centering
    \includegraphics[width=\columnwidth]{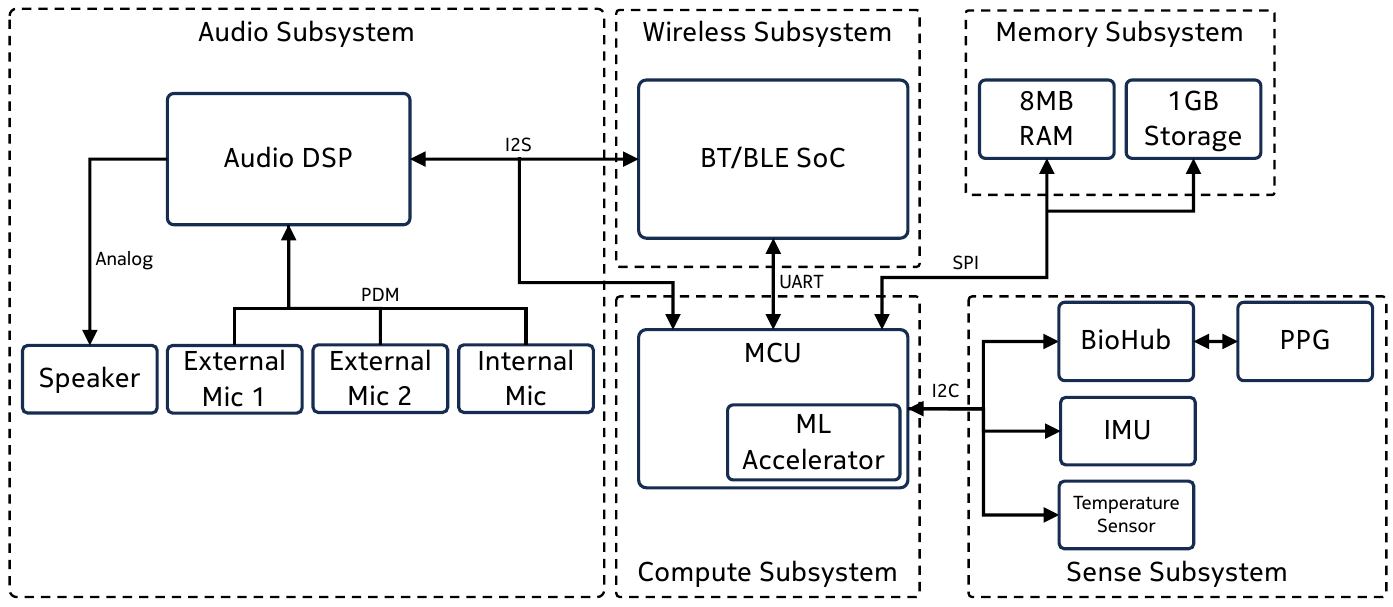}
    \vspace{-5mm}
    \caption{OmniBuds Hardware Block Diagram.}
    \label{fig:hardwarearchitecture}
\end{figure}

\begin{figure}
    \centering
    \includegraphics[width=0.8\columnwidth]{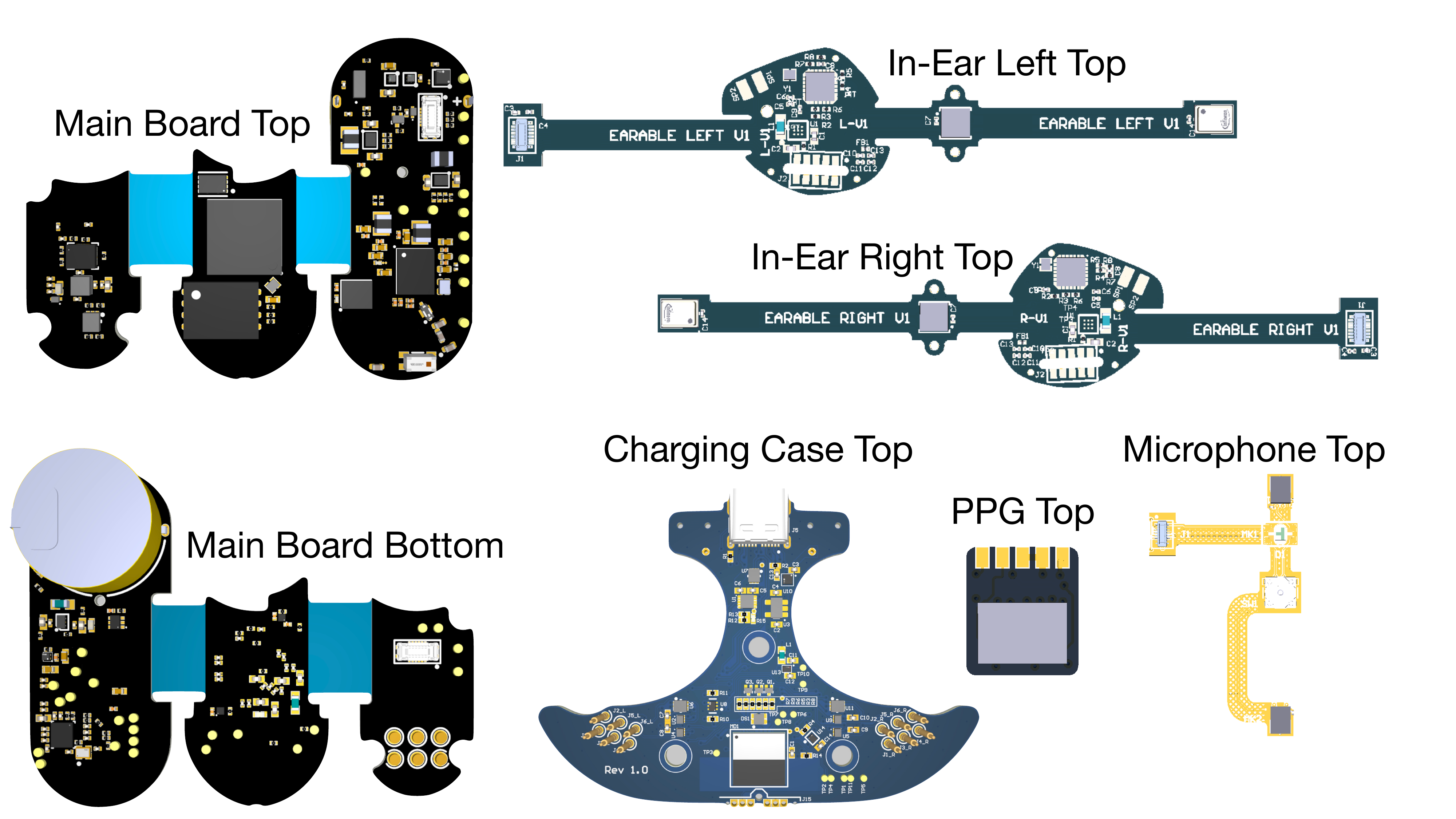}
    \vspace{-3mm}
    \caption{OmniBuds PCBs.}
    \label{fig:pcbs}
    \vspace{-5mm}
\end{figure}

OmniBuds are fully functional True Wireless Stereo (TWS) earbuds with standard features like music playback, calls, Active Noise Cancellation (ANC), and Acoustic Transparency (or Pass-through). What distinguishes them is the integration of additional sensors and computational units, transforming them into a powerful platform for earable computing research. Both earbuds share identical hardware, enabling multi-device computation and sensing. This symmetry allows computational and sensing tasks to be distributed between the two earbuds, such as one performing signals pre-processing while the other handles intensive machine learning tasks, or both working together for enhanced spatial sensing.

At the core of OmniBuds hardware (Figure~\ref{fig:hardwarearchitecture}) is a dual-core microcontroller (MCU) with a CNN accelerator for efficient on-device machine learning. A dedicated audio DSP handles audio processing, and the BioHub processor manages PPG data for vital signs estimation, all designed to emphasise local computation for real-time data processing and machine learning.

OmniBuds are equipped with a comprehensive suite of sensors selected for their low-power characteristics and advanced functionalities: a 9-axis IMU, a 3-wavelength PPG sensor, a medical-grade temperature sensor, and three microphones. Each earbud includes 1GB of non-volatile memory and 8MB of RAM, offering ample capacity for complex applications, such as large datasets and sophisticated machine learning models. For interaction, the earbuds feature a button and an RGB LED. Figure~\ref{fig:pcbs} shows the OmniBuds PCBs, consisting of a mix of rigid, flex-rigid, and flex designs.

The earbuds are powered by a 105mAh battery, providing up to 6 hours of music playback and approximately 8 hours of PPG sensing. The charging case, featuring a 630mAh battery, allows for multiple recharges of the earbuds, extending their total usage time.

In the following sections, we first analyse the OmniBuds form factor and the positioning of the sensors before delving into the main HW submodules: computation, communication and sensing.

\subsection{Form Factor and Sensor Placement}
\label{sec:hw_form_factor}

\begin{figure}
    \centering
    \includegraphics[width=0.8\columnwidth, keepaspectratio]{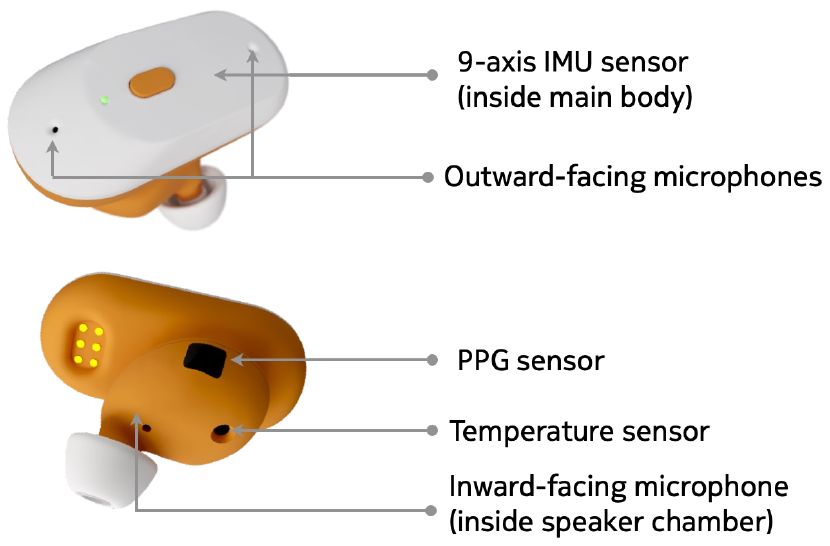}
    \vspace{-2mm}
    \caption{Placement of the sensors in OmniBuds.}
    \label{fig:OmnibudImage}
    \vspace{-5mm}
\end{figure}

Form factor and ergonomics design are crucial in wearables, impacting usability, performance, and sensor data quality. When designing OmniBuds, we considered factors affecting comfort, sensor accuracy, and manufacturing ease. Enclosed in an injection-moulded plastic cover, OmniBuds measure 4.3 x 3 x 2.2 cm and weigh around 12g,
roughly twice as much as other commercial true wireless earbuds (e.g., Apple AirPods Pro 2: 5.3 grams~\cite{appleAirpodsWeight}, Bose Quite Comfort Earbuds II: 6.24 grams~\cite{boseindiaBoseQuietComfort})
Despite this, OmniBuds remain comfortable even when worn for extended periods. 
OmniBuds feature a more capable and intricate sensor suite than comparable earbuds, requiring a larger battery. Specifically, the battery accounts for about 20\% of the earbud weight (approximately 2.4 grams).

Sensor placement is a critical element of the OmniBuds design, where performance, user comfort, and manufacturing feasibility are carefully balanced. Each positioning decision is driven by extensive validation studies and the need to optimise the quality of data collected without compromising wearability or ease of production. Figure~\ref{fig:OmnibudImage} illustrates the placement of the various sensors within the OmniBuds.

The arrangement of the three microphones is particularly strategic. Two outward-facing microphones, positioned at the top and bottom of each earbud, facilitate essential acoustic features such as ANC, Transparency mode, and beamforming for voice capture. The positioning allows the microphones to target sounds from the user's environment while enabling effective noise cancellation by capturing ambient noise from multiple angles. Meanwhile, the inward-facing microphone, located within the speaker chamber, serves multiple purposes. While its primary role is as a reference microphone to enhance ANC performance, it also acts as a sensitive detector for internal body sounds, as explored in many studies~\cite{ferlini2021eargate, ma2021oesense, truong2022non}. To maximize the accuracy of in-ear recordings, the earbuds are designed to ensure a snug fit in the ear canal, exploiting the occlusion effect for higher-quality body sound detection. The combination of interchangeable ear-tips and an ergonomic earbud shape helps achieve this fit, ensuring both acoustic performance and user comfort.

The placement of the 9-axis Inertial Measurement Unit (IMU) is equally deliberate. Positioned centrally on the outer plastic cover of the earbuds, the IMU is optimally located to capture a broad range of movements, from subtle facial gestures to larger head and body movements
supporting diverse applications that rely on motion data.

The photoplethysmography sensor (PPG) is crucial in OmniBuds since it supports the measurement of all vital signs (except temperature) directly or indirectly. The accuracy of PPG-derived metrics depends on several factors, including sensor location. In designing OmniBuds, we conducted a systematic validation study to determine the optimal placement for a PPG sensor in an earable~\cite{ferlini2021ear}. We explored three placements: concha, ear canal, and behind the auricle (or pinna). Our study concluded that the best location for the PPG sensor in OmniBuds is facing towards the concha. Although in-ear canal placement provides slightly more accurate data, the concha placement preserves a higher level of user comfort and practicality for everyday use, which are key factors for long-term wearability in earables.

Finally, the location of the temperature sensor is determined by the need to measure body core temperature with high accuracy, while accounting for usability and manufacturing constraints. While placing the sensor in the ear canal would provide the most accurate readings~\cite{ferlini2021ear}, it is instead placed against the skin of the concha, which still offers a reliable proxy for body core temperature. This location strikes the ideal balance between accuracy and manufacturing feasibility, allowing for effective temperature sensing without requiring overly complex computational corrections.

In summary, the positioning of sensors in OmniBuds reflects a careful evaluation of trade-offs, balancing scientific rigour with practical considerations to ensure the best possible user experience and data quality.

\subsection{Compute Subsystem}
\label{sec:hw_computation}

\subsubsection{Main Processor and CNN Accelerator}
At the core of the OmniBuds' computation system is a low-power dual-core MCU (Cortex-M4F and RISC-V) with a floating point unit, providing robust processing power. The MCU features 512KB of Flash memory and 128KB of SRAM, making it well-suited for managing real-time tasks and coordinating the various sensors and subsystems. 

To further enhance its computational capabilities, OmniBuds integrates a dedicated hardware-based Convolutional Neural Network (CNN) accelerator. The accelerator consists of 64 convolutional processors, each equipped with its own pooling engine, input cache, weight memory (from 8-bit down to 1-bit width), and convolution engine, enabling highly parallelised execution of machine learning models. The CNN accelerator supports up to 64 layers and can handle a maximum input size of 1024x1024, allowing it to execute a wide range of neural network architectures, including feed-forward models, residual networks, recurrent models and encoder-decoder designs. 

\subsubsection{Specialised Computational Units}
In addition to the main processor and CNN accelerator, OmniBuds include specialised components designed for task-specific workloads, enhancing the platform’s ability to efficiently handle diverse data types while maintaining power efficiency.

One such component is the \textit{Audio DSP}, which manages audio pipelines and processing tasks. It consists of two sub-cores: the Fast-DSP core for low-latency applications like ANC, and the Slow-DSP core for less time-critical tasks. The DSP allows efficient audio processing without burdening the main MCU, and its flexibility enables it to support a variety of tasks.

Another key component is the \textit{BioHub}, a co-processor dedicated to biosignal processing. The BioHub manages the PPG sensor and its dedicated 6-axis IMU, computing the user’s vitals in real-time while minimising power consumption. By offloading tasks from the main MCU, the BioHub enables parallel execution, allowing the main processor to enter a low-power state when idle, optimising energy use.

\subsection{Wireless Subsystem}
\label{sec:hw_communication}
OmniBuds employ a dedicated Bluetooth System-on-Chip (SoC) for wireless communication.
This SoC supports both Bluetooth Classic for audio streaming and Bluetooth Low Energy (BLE) for data communication, ensuring seamless connectivity with external devices. LE Audio can also be supported with software updates.
In Bluetooth Classic mode, the SoC manages communication with the host device, notifying the main MCU of events like connections and audio stream changes. For BLE, it provides a transparent API for data transmission, allowing the main processor to focus on other tasks without handling communication protocols.

This division of responsibilities between the Bluetooth SoC and main MCU ensures that OmniBuds can maintain efficient wireless communication while keeping overall power consumption low. We will delve further into the software management of this communication system in Section~\ref{sec:sw_comms}.

\subsection{Sensing Subsystem}
\label{sec:hw_sensors}

\subsubsection{Photoplethysmography sensor (PPG)}

\begin{figure}
    \vspace{-2mm}
    \centering
    \includegraphics[width=0.85\columnwidth]{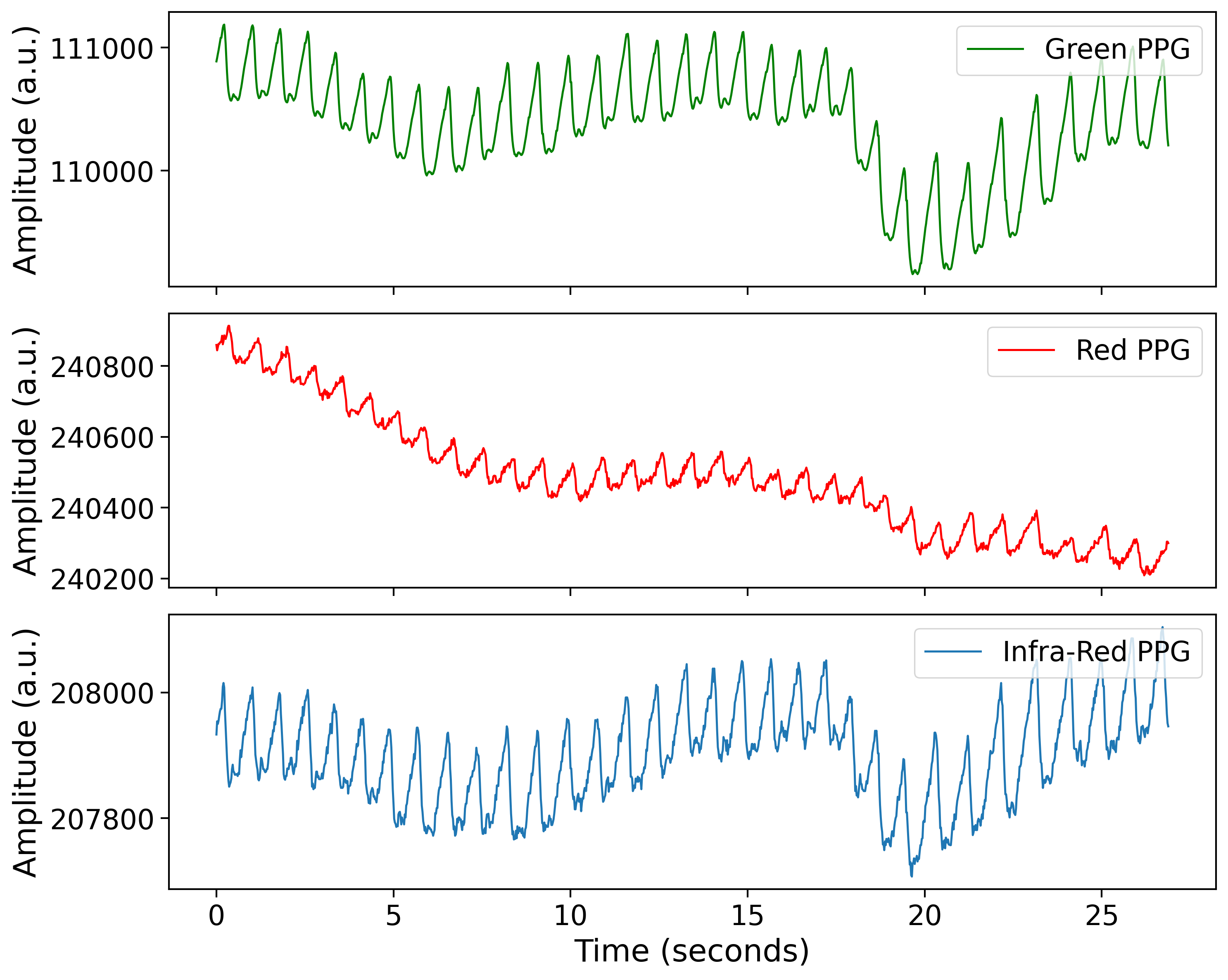}
    \vspace{-5mm}
    \caption{PPG data sampled at 50Hz during rest.}
    \label{fig:ppg_sample}
    \vspace{-2mm}
\end{figure}

PPG is a non-invasive optical technique that detects blood volume changes by measuring light absorption. It uses an LED to emit light through the skin and a photodiode to measure the reflected light. As blood volume increases, more light is absorbed; as it decreases, less light is absorbed. This data provides insights into physiological parameters like heart rate, respiration rate, blood pressure, and oxygen saturation~\cite{ferlini2021ear, balaji2023stereo, romero2024optibreathe}.

One of the criteria that drove the choice of the specific PPG sensor used in the OmniBuds, is its flexibility and the number of different configurations/parameters that it supports.
The sensor includes three LEDs which generate three different wavelengths, green (530nm), red (660nm) and infrared (880nm), and a single photodiode with peak sensitivity at 860nm and a spectral bandwidth range from 420nm to 1020nm. 
Figure~\ref{fig:ppg_sample} shows an example of 50Hz PPG data collected from the OmniBuds at the three available wavelengths where the vascular pulses and breathing-related modulations are clearly visible.

\subsubsection{Inertial Measurement Unit}

\begin{figure}
    \vspace{-2mm}
    \centering
    \includegraphics[width=0.85\columnwidth]{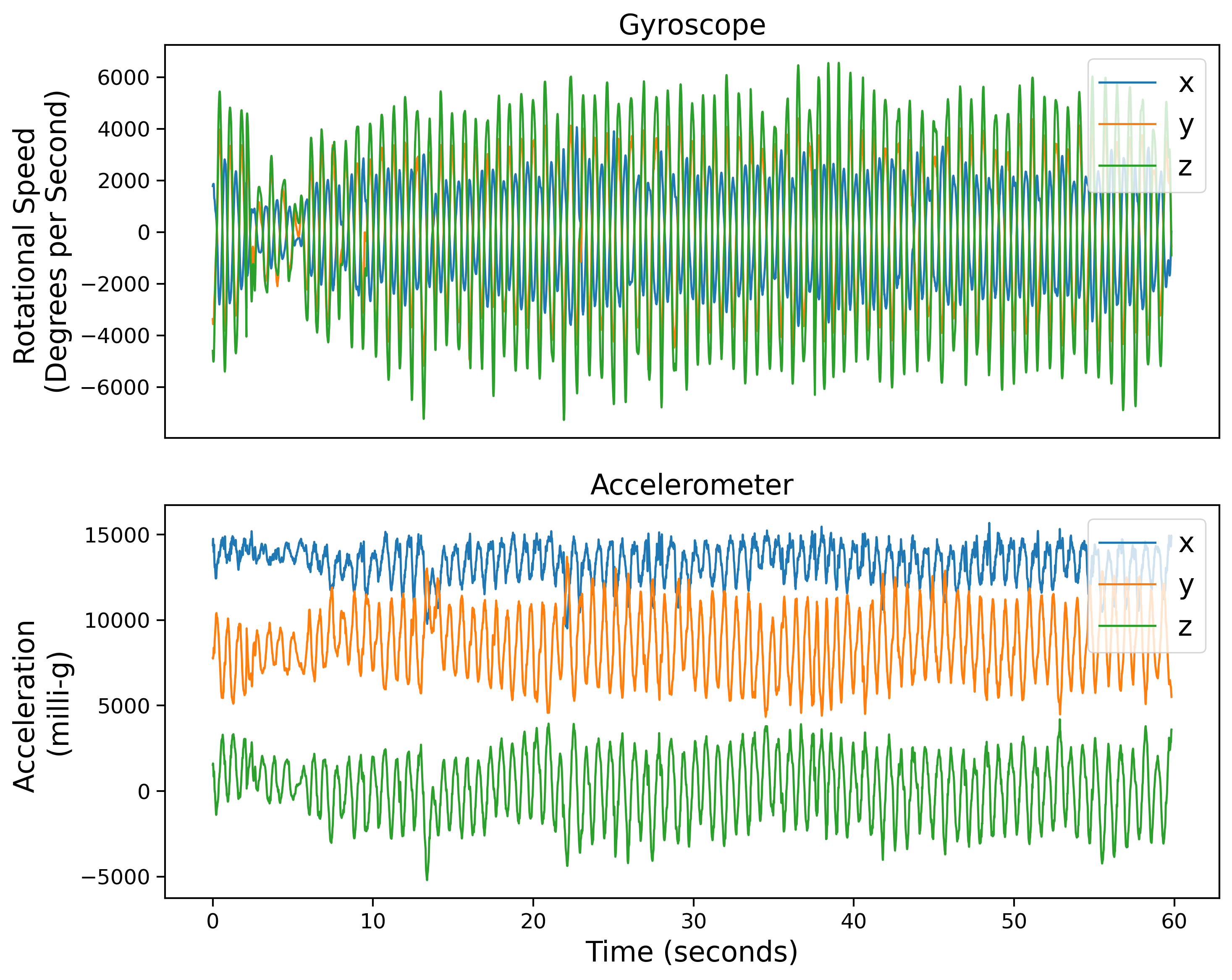}
    \vspace{-5mm}
    \caption{IMU data sampled at 100Hz during head nodding.}
    \label{fig:imu_sample}
    \vspace{-5mm}
\end{figure}

The OmniBuds feature two separate Inertial Measurement Units (IMUs). The first one is a 6-axis IMU (accelerometer and gyroscope) which is co-located on the same PCB that hosts the PPG sensor and is dedicated to PPG motion artefact removal.
In addition to that, OmniBuds also features a 9-axis IMU (i.e., accelerometer, gyroscope, and magnetometer). Whilst this can be used to track macro motions such as walking and running, it can also be exploited by researchers to perform head motion tracking and 3D pose estimation~\cite{ferlini2019head, lee2019automatic}.

The 9-axis IMU is a modern low-power component which offers natively several features like step detection, step counting, device orientation and single and double click detection. More importantly, the IMU allows the execution of simple on-chip ML models courtesy of its embedded Machine Learning Core module (MLC).
The MLC works directly by taking as input the data streams coming from accelerometer, gyroscope and magnetometer, computing features on the data and passing these features to user-defined decision tree models. This enables users to run simple IMU-based inference tasks directly on the sensor waking up the main MCU only when an inference is available. This results in great power saving compared to waking up every time a new raw data sample is available and computing the inference on the MCU.
Figure~\ref{fig:imu_sample} shows data collected from the main OmniBuds' IMU while the user was nodding.

\subsubsection{Temperature}
Temperature sensing is achieved using an infrared sensor directed at the concha area of the ear, allowing for applications like body core temperature and fertility tracking. 

The sensor was selected for its high accuracy of $\pm 0.2^\circ$ C within the range of $35^\circ$ C to $42^\circ$ C\footnote{Ambient temperature between $15^\circ$ C and $40^\circ$ C.}, which is ideal for human sensing applications. The total measurement range spans from $-20^\circ$ C to $100^\circ$ C, making it versatile for various conditions. Additionally, the sensor’s configurable sampling rate of up to $64$~Hz allows it to detect minute and quick temperature fluctuations, providing enhanced precision in continuous monitoring scenarios.

\subsubsection{Microphones}

\begin{figure}
    \centering
    \includegraphics[width=\columnwidth]{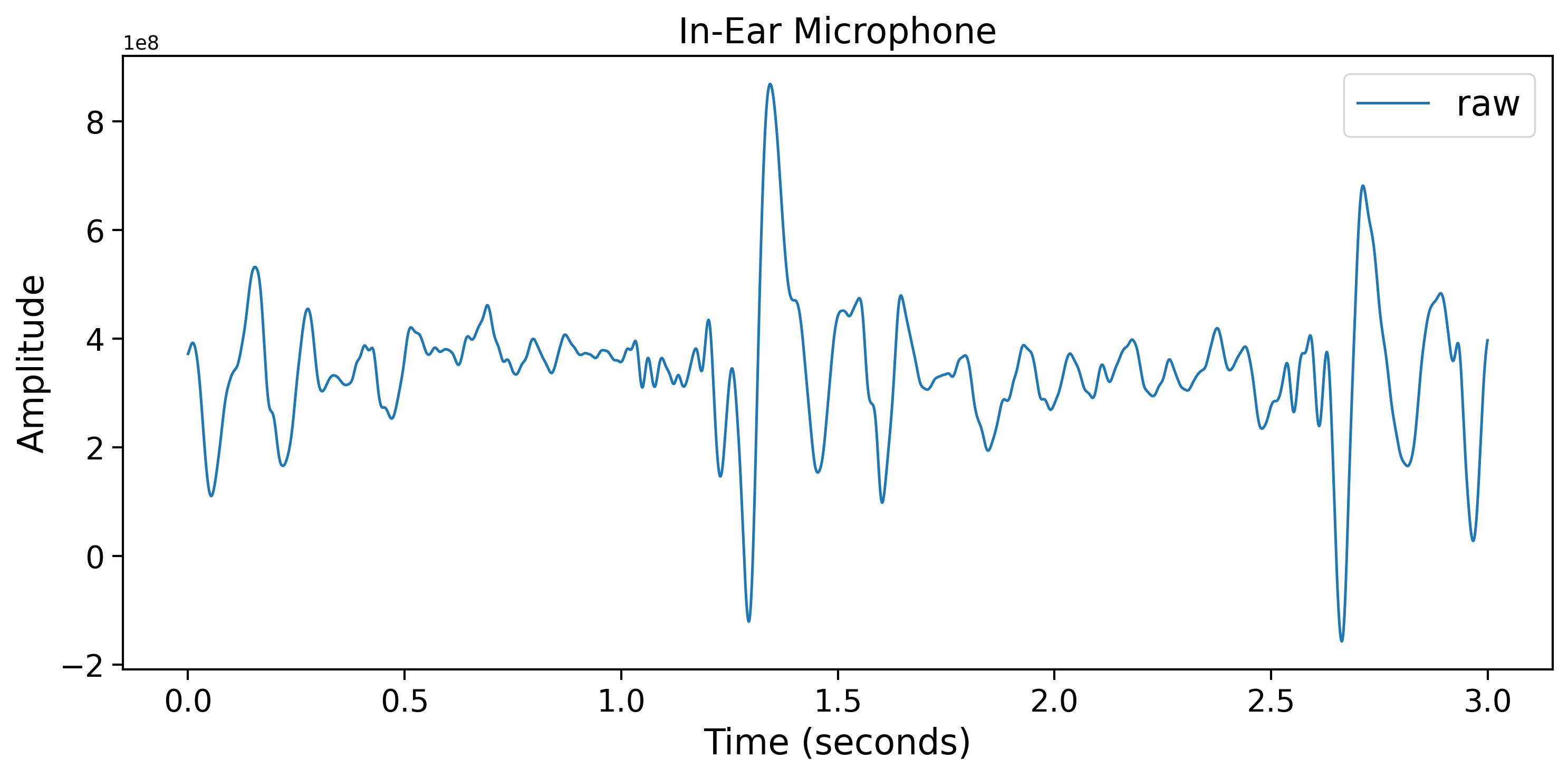}
    \vspace{-8mm}
    \caption{Internal microphone data sampled at 48kHz during silence.}
    \label{fig:in_ear_mic_sample}
\end{figure}

The OmniBuds come equipped with three microphones. Two are facing outward while one is embedded in the speaker chamber and facing the user's ear canal, as discussed in section~\ref{sec:hw_form_factor}. When selecting microphones for OmniBuds, we carefully balanced the requirements of standard acoustic features like active noise cancellation and transparency mode with the need for high-fidelity sensing performance. Microphones optimised for ANC and Transparency must capture environmental sounds accurately, while sensing applications require them to detect subtle internal body sounds, such as heartbeats. Ensuring that the microphones could excel at both tasks was critical to the design. We selected components that offered a sufficient dynamic range and sensitivity to perform well in both acoustic and sensing contexts, acknowledging the inherent trade-offs in optimising for dual functionality.

Figure~\ref{fig:in_ear_mic_sample} shows data captured from the OmniBuds' internal microphone at 48kHz during silence. The signal clearly shows how the microphone is capable of capturing heart sounds which are used for diverse applications, like heart rate measurement and blood pressure estimation~\cite{truong2022non, butkow2023heart}.

\section{Software Architecture}
\label{sec:software_architecture}

\begin{figure}
    \centering
    \includegraphics[width=\columnwidth]{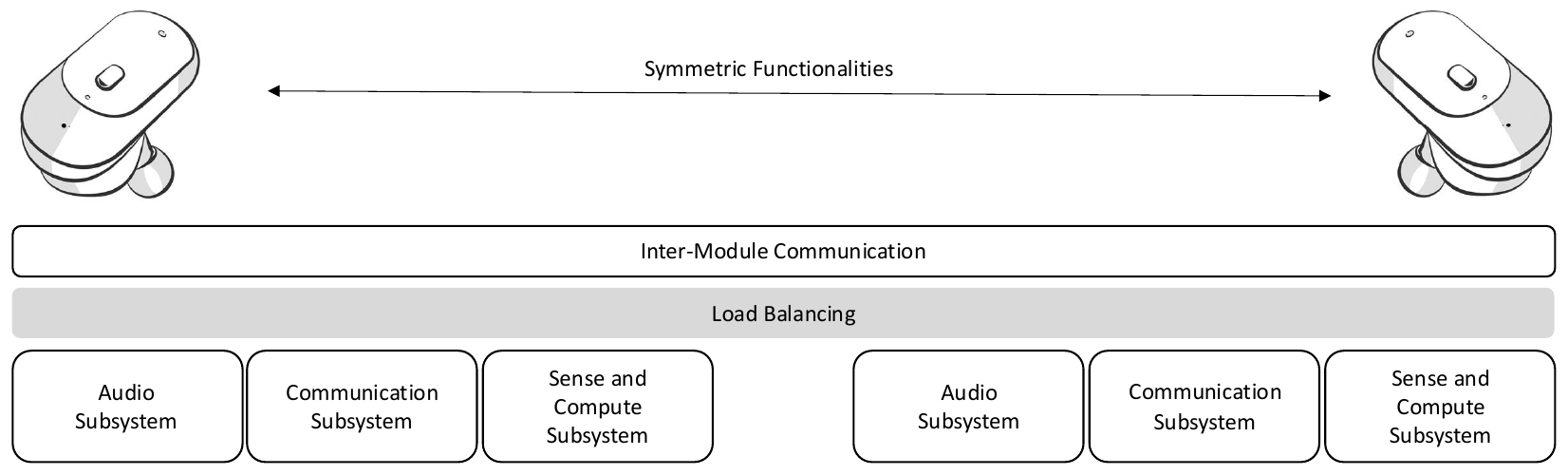}
    \vspace{-5mm}
    \caption{OmniBuds Ecosystem and SW Overview.}
    \label{fig:obconnection}
    \vspace{-5mm}
\end{figure}

\begin{figure}[t]
    \centering
    \includegraphics[width=\columnwidth]{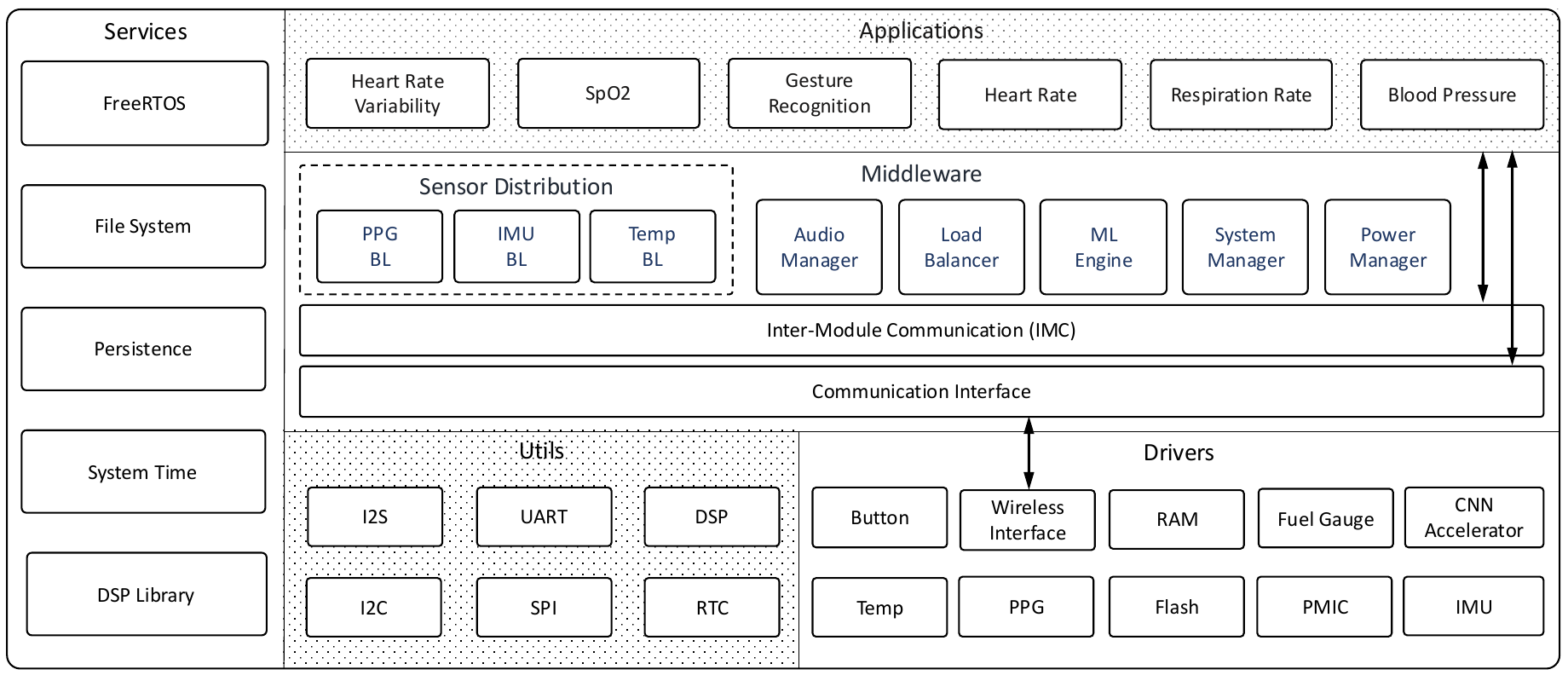}
    \vspace{-8mm}
    \caption{OmniBuds Software Architecture.}
    \label{fig:sw_architecture}
    \vspace{-3mm}
\end{figure}

The OmniBuds ecosystem consists primarily of two earbuds with identical hardware, as discussed earlier, and is complemented by a charging case and an optional host device connected via Bluetooth Classic (BT) and/or Bluetooth Low Energy (BLE). Figure~\ref{fig:obconnection} provides an overview of the OmniBuds ecosystem. The software architecture is designed to fully leverage this multi-device setup, enabling efficient coordination across devices.

The software architecture of OmniBuds adopts a layered design that allows developers to extend the platform’s functionality while respecting the hardware's constraints. As shown in Figure~\ref{fig:sw_architecture}, the architecture consists of four key layers:

\begin{itemize}
    \item \textbf{Hardware Abstraction Layer (HAL):} Encapsulates device drivers and utility libraries, ensuring portability across different hardware configurations.
    
    \item \textbf{Middleware Layer:} Provides essential APIs for applications to communicate with and efficiently use hardware resources. It ensures independent applications can run without conflicts by managing access to hardware in a modular manner.
    
    \item \textbf{Applications Layer:} Hosts algorithms that implement functionalities such as vital signs monitoring, leveraging localised computing to optimise performance and conserve power.
    
    \item \textbf{Services Layer:} Provides system-wide services such as file management, persistent storage, and task scheduling. 
    The platform runs FreeRTOS which is packaged as part of the services. The applications are free to use FreeRTOS primitives for their own purposes, but should use the OmniBuds middleware APIs for the functionalities provided by the platform and to access the hardware infrastructure.
\end{itemize}

This layered design ensures flexibility, scalability, and ease of development while maintaining tight integration between software and hardware. The following sections focus on three critical subsystems: communication, sense and compute, and load balancing, each playing a vital role in the platform's efficient operation.

\subsection{Communication Subsystem}
\label{sec:sw_comms}

The Communication Subsystem in OmniBuds manages data exchange within the platform and with external devices. It consists of two main components: Inter-Module Communication (IMC) for internal communication between software modules and the Communication Interface, which uses the BLE protocol to connect with external peripherals (Figure~\ref{fig:sw_architecture}).

\subsubsection{Inter-Module Communication (IMC)}

IMC facilitates seamless and transparent communication between software modules, both locally and across devices in the OmniBuds ecosystem. This approach ensures that software modules can interact without needing to manage the complexities of local versus remote communication.

This system is based on the observer pattern~\cite{gamma1995design}, where components register for updates from extendable triggers. When a trigger is activated, the registered modules receive notifications with the relevant data. IMC messages are asynchronous and can carry up to 64 bytes of data.

The IMC API offers flexibility in message dissemination, allowing modules to specify if a message is sent to the local device, a peer device (i.e. the other earbud), other devices in the OmniBuds ecosystem, or all of the above. The underlying IMC logic handles message queuing efficiently, even when devices use different physical layer connections. The APIs for sending and receiving IMC messages are shown below:

\begin{lstlisting}[basicstyle=\small, language=C, keywordstyle=\color{blue}, commentstyle=\color{gray}, morekeywords={uint8_t, uint16_t, uint32_t, triggerEvtCB_t}, emph={ob_IMCSendMessage, ob_IMCRegisterMessageCallback},emphstyle=\color{olive}]
void ob_IMCSendMessage(uint8_t triggerID, uint8_t *dataBuffer, uint8_t dataLength, uint8_t destination)

int ob_IMCRegisterMessageCallback(uint8_t triggerID, triggerEvtCB_t *triggerCB)
\end{lstlisting}

\subsubsection{Communication Interface}
\label{sec:comm_external}

The Communication Interface module in OmniBuds manages data exchange with external devices via Bluetooth Low Energy (BLE) and handles audio playback and phone calls through Bluetooth Classic. 

To external devices, the two OmniBuds earbuds appear as a single device, with the \textit{primary} earbud managing the connection and forwarding messages to the \textit{secondary} earbud. This architecture simplifies communication while maintaining seamless interaction between both earbuds. Details on how this supports load balancing and battery optimisation are discussed in Section~\ref{sec:load_balancer}.

For data exchange, OmniBuds introduces \textit{addressable peripherals}, which can be either physical sensors (e.g., IMU or PPG) or virtual peripherals (e.g., heart rate derived from PPG). These allow third-party devices to send commands, collect data, and interact with specific peripherals via a consistent interface. BLE communication is encrypted to ensure data privacy and security.

BLE communication is based on three message types:
\begin{itemize}
    \item \textbf{Data messages:} For large or continuous data streams.
    \item \textbf{Event messages:} For handling sporadic peripheral triggers.
    \item \textbf{Configuration messages:} For configuring peripheral settings such as sampling rate or power mode (using different endpoints associated with each peripheral).
\end{itemize}

The main APIs for managing BLE communication are as follows:
\begin{lstlisting}[basicstyle=\small, language=C, keywordstyle=\color{blue}, commentstyle=\color{gray}, morekeywords={uint8_t, uint16_t, uint32_t, triggerEvtCB_t, ob_periphID_t, rxCommEvtCB, ob_msgErrorCode_t}, emph={ob_registerPeripheralCallback, ob_sendDataMessage, ob_sendEventMessage, ob_sendConfigMessage, ob_sendConfigMessageResponse},emphstyle=\color{olive}]
// Register callback to receive data/event/config messages
int ob_registerPeripheralCallback(ob_periphID_t peripheralD, rxCommEvtCB *callbac);

// Send a data/event message
void ob_sendDataMessage(ob_periphID_t peripheralD, uint8_t *data_buf, uint16_t data_len);
void ob_sendEventMessage(ob_periphID_t peripheralD, uint8_t *data_buf, uint16_t data_len);

// Send a configuration message
void ob_sendConfigMessage(ob_periphID_t peripheralD, uint8_t configEndpoint, uint8_t *data_buf, uint16_t data_len);

// Send a configuration message response
void ob_sendConfigMessageResponse(ob_periphID_t peripheralD, uint8_t configEndpoint, ob_msgErrorCode_t errCode, uint8_t *data_buf, uint16_t data_len);
\end{lstlisting}

OmniBuds also manages audio playback and phone calls via Bluetooth Classic, using the Hands-Free Profile (HFP) for calls and Audio-Video Remote Control Profile (A2DP) for playback, coordinated by the communication subsystem and wireless chipset.

\subsection{Sense and Compute Subsystem}
\label{subsec:processingandsensingsubsystem}

The Sense and Compute Subsystem in OmniBuds provides applications with primitives to manage the platform's computational units and sensors efficiently. Three critical modules define this subsystem: the Sensor Distribution Module, which manages sensor data access for multiple applications; the Machine Learning Engine, responsible for executing models on the CNN accelerator; and the Audio Manager, which oversees audio pipelines within the dedicated Audio DSP. The following sections provide an overview of these key modules.

\subsubsection{Sensor Distribution Module}

The Sensor Distribution Module simplifies access to sensor data for multiple applications, particularly for sensors like the IMU, PPG and temperature. It allows application developers to easily integrate sensor data without managing conflicts or sensor resource issues. 

Different Business Logic modules are provided for the available sensors (Figure~\ref{fig:sw_architecture}) which are used as singleton~\cite{gamma1995design} from multiple applications. Applications register with the module by specifying their requirements for the sensor data, such as sampling rate, window length, and other parameters. The module then configures the sensor accordingly and efficiently delivers the requested data to each application. To further enhance efficiency, the module leverages on-chip computational features, such as the BioHub and the machine learning core in the IMU, to process data locally when necessary. This reduces the computational load on the main MCU, conserves power, and ensures that sensor data is processed and delivered in a timely manner.

\subsubsection{Machine Learning Engine}

The Machine Learning Engine manages the execution of machine learning models on the CNN accelerator, ensuring efficient on-device inference. The use of this module involves a complete pipeline, from offline model training and synthesis to deployment and execution, ensuring compatibility with the accelerator’s architecture.

Models are trained and optimised offline before being transferred to the OmniBuds via BLE. The engine then loads and executes the model on the CNN accelerator, making inference results available to other applications. To conserve power, the engine dynamically configures the accelerator, shutting down unused hardware sections during inference.

Looking ahead, the Machine Learning Engine is designed to support future capabilities such as splitting inference tasks across both earbuds, enabling distributed computation, and running multiple models concurrently. These features will extend the flexibility of the system, enhancing performance and energy efficiency.

\subsubsection{Audio Manager}

\begin{figure}
    \centering
    \includegraphics[width=\columnwidth]{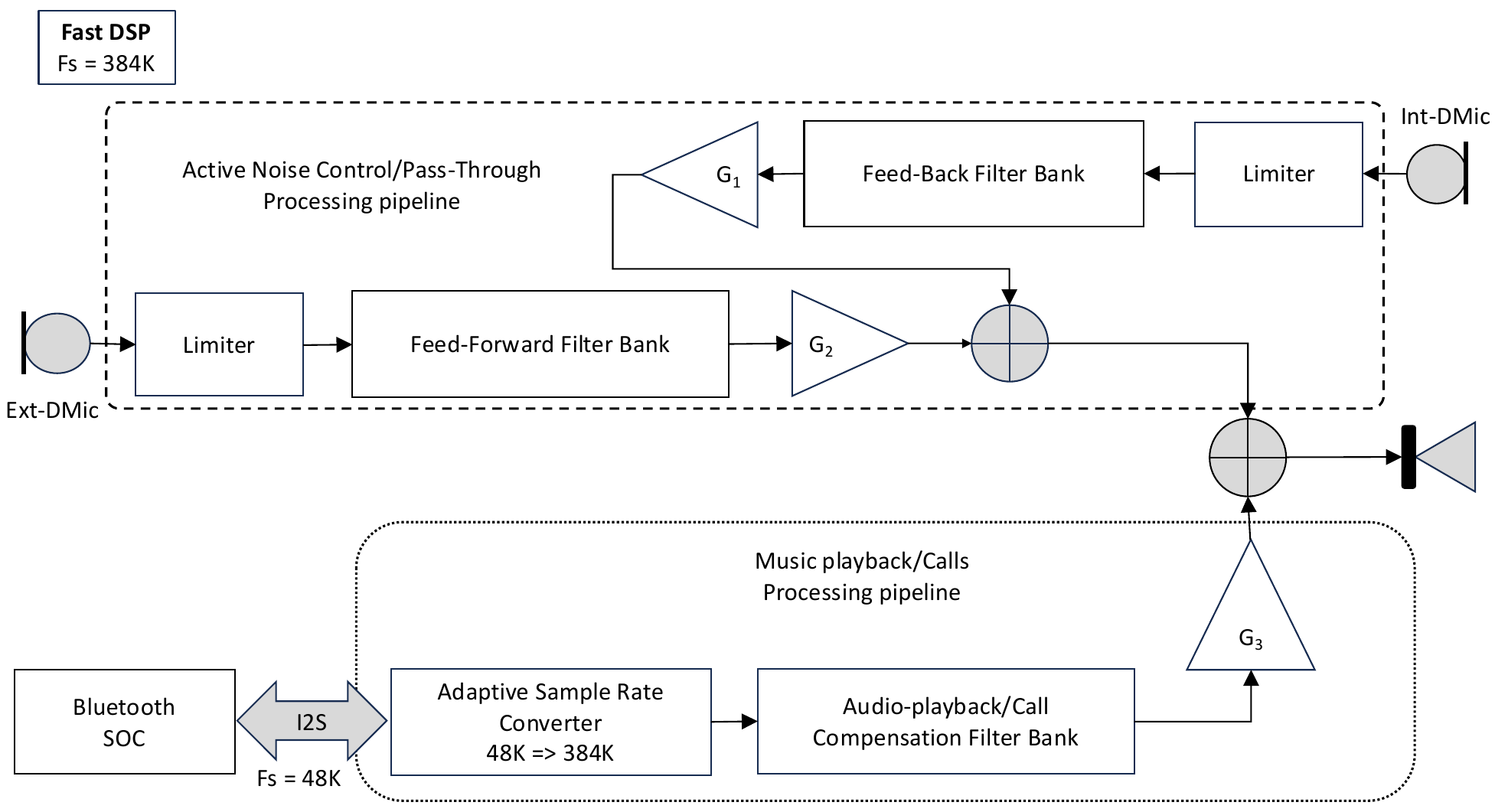}
    \vspace{-5mm}
    \caption{Fast DSP algorithms for ANC/Pass-through and music playback and calls.}
    \label{fig:fastdsp_algorithms}
\end{figure}

\begin{figure}
    \centering
    \includegraphics[width=\columnwidth]{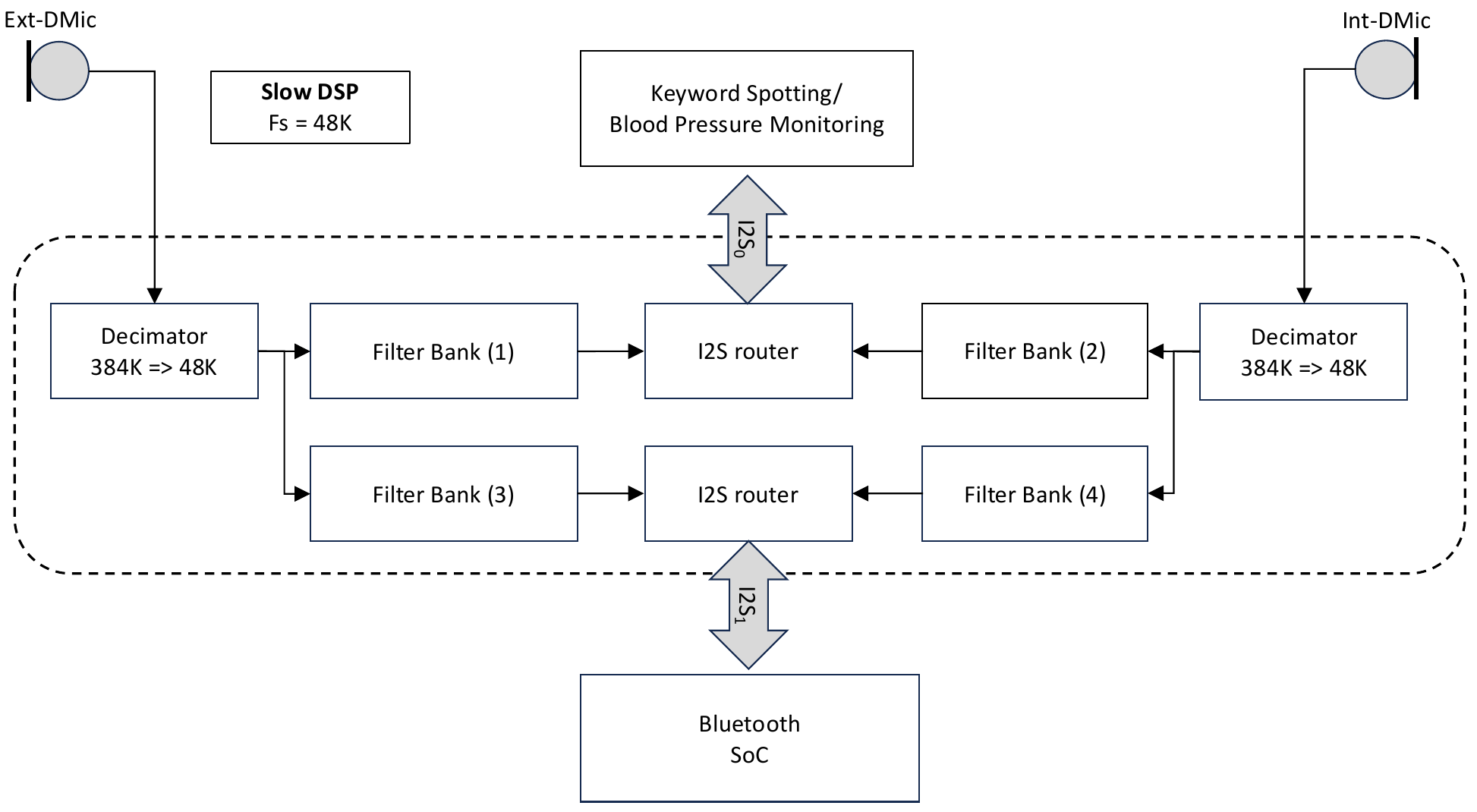}
    \vspace{-5mm}
    \caption{Slow DSP algorithms to support sensing applications like spoken keyword spotting and multi-modal blood pressure estimation.}
    \label{fig:slowdsp_algorithms}
    \vspace{-5mm}
\end{figure}

The Audio Manager handles the audio processing pipelines on the dedicated audio DSP, divided into \textit{time-critical} and \textit{non-time-critical} categories.

\paragraph{Time-Critical Pipelines}
Time-critical pipelines include processes such as Active Noise Cancellation (ANC) and pass-through functionality. These pipelines run on the Fast-DSP core, which is optimised for low-latency processing, operating at a high sampling frequency of $F_s = 384$ KHz. As shown in Figure~\ref{fig:fastdsp_algorithms}, the Fast-DSP processes both ANC and pass-through in a shared structure, with differences in the parameters loaded into the feed-forward and feedback filter banks. For safety, when the pass-through functionality is enabled, the limiter ensures that no sound pressure levels exceed 85 dB-SPL. The Fast-DSP also manages music playback and phone calls, which are processed in a similar pipeline to ANC and pass-through but differ in the filter parameters used. In both music and phone calls, audio content coming from the Bluetooth SOC needs to be routed to the earbud's speaker to be played back.

While ANC and pass-through are mutually exclusive, they can coexist with music playback or calls, where the outputs from both pipelines are mixed before being sent to the speaker. In this dual-processing configuration, different filter parameters are loaded to compensate for the effect of ANC or pass-through on the ear canal’s acoustics.

\paragraph{Non-Time-Critical Pipelines}
Non-time-critical pipelines (see Figure~\ref{fig:slowdsp_algorithms}) handle tasks that are less latency-sensitive, such as keyword spotting, blood pressure monitoring, and heart rate detection, running on the Slow-DSP core at a lower sampling frequency of $F_s = 48$ KHz. To achieve this, MEMS microphone signals—originally sampled at higher frequencies—are decimated.

The Slow-DSP offloads audio-related tasks, such as sensor sampling, filtering, and signal pre-conditioning, reducing the load on the MCU. For tasks like keyword spotting or blood pressure monitoring, the DSP routes MEMS microphone signals to the microcontroller via the $I2S_0$ bus, where further processing or inference (using the CNN accelerator) takes place. During calls, the DSP similarly routes MEMS microphone signals to the Bluetooth SoC, ensuring efficient transmission to the host device while the MCU remains focused on higher-level application logic.

\subsection{Load Balancing Subsystem}
\label{sec:load_balancer}

OmniBuds prioritise efficient battery management and balanced system performance given the limited battery capacity. The Load Balancing Subsystem ensures that both communication and computational tasks are distributed optimally across the two earbuds, enhancing battery life.

In terms of communication, OmniBuds present themselves as a single device to external devices (Section~\ref{sec:comm_external}). The primary earbud maintains the connection with the external device, routing any necessary information to and from the secondary earbud. To prevent one earbud from being overly burdened, the system periodically alternates the roles of primary and secondary between the two earbuds, balancing the communication load and distributing battery usage evenly.

For computational load balancing, OmniBuds include a dedicated Load Balancing Module. This module autonomously determines, at runtime, which earbud should execute a given task. 
The Load Balancer operates as a system-wide component, running on both earbuds and maintaining a shared database of peripherals and their states. When a peripheral is enabled, the Load Balancer determines its operational state based on the active policies, ensuring that tasks are executed efficiently across the earbuds.

\section{Smartphone Application}
\label{sec:mobile_app}

\begin{figure}
    \centering
    \includegraphics[width=0.5\columnwidth]{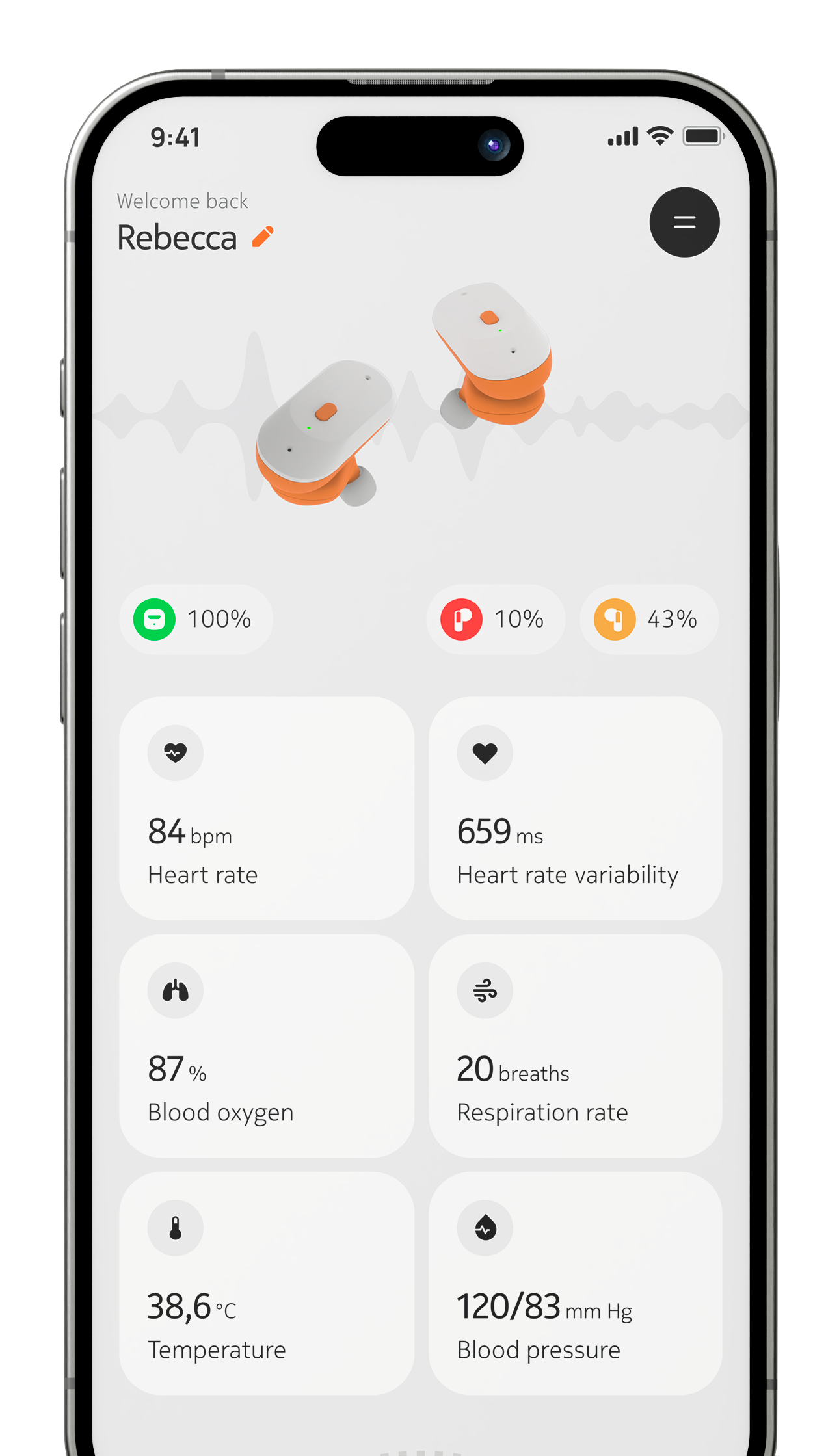}
    \vspace{-2mm}
    \caption{Dashboard of the OmniBuds companion app.}
    \label{fig:ob_app_dashboard}
    \vspace{-5mm}
\end{figure}

The OmniBuds come with a companion app for both iOS and Android, providing real-time monitoring and visualisation of physiological data. Through a dashboard (Figure~\ref{fig:ob_app_dashboard}), users can view six vital signs: heart rate, heart rate variability, respiratory rate, blood oxygen saturation, body temperature, and blood pressure. Historical data is accessible with a simple tap, enabling health tracking over daily, monthly, and yearly intervals. The app also allows users to control acoustic settings, switch between transparency and noise-cancelling modes, and initiate blood pressure measurements directly from the dashboard.

For advanced users and developers, the app includes a Developer Mode, which allows the collection of raw, unprocessed sensor data. The data can be saved as \textit{.csv} files for further analysis, offering invaluable opportunities for experimentation and research. Additionally, the Developer Mode enables modifications to parameters such as sensor sampling rates, LED current for the PPG sensor, and the angular rate for the accelerometer, among others, as well as the choice of which sensor to enable/disable. In addition, when data is being recorded, users can visualise real-time plots of raw sensor data from a dedicated Plot View. This allows for immediate feedback on parameter fine-tuning.

\section{Applications of OmniBuds}

OmniBuds' unique combination of sensors, computational capabilities, and compact form factor opens the door to a wide range of applications across multiple fields. While some of these applications have already been implemented, others remain potential future use cases. The following examples, although not exhaustive, showcase the versatility of OmniBuds, highlighting how the platform’s advanced features can be leveraged for innovative research and practical solutions in earable computing.

\subsection{Existing Applications}

\subsubsection{Vital Signs Monitoring}

The advanced sensing capabilities of OmniBuds, coupled with on-device processing and efficient use of computational resources, make it a versatile platform for continuous vital signs monitoring. By integrating multiple sensor modalities in a compact form factor, OmniBuds enable precise, real-time health tracking, supporting a wide range of research and healthcare applications. Out-of-the-box OmniBuds can monitor heart rate (HR), heart rate variability (HRV), respiration rate (RR), blood oxygen saturation (SpO$_{2}$), blood pressure, and skin temperature. While skin temperature is measured directly, the other vitals are computed on-device, utilising either the main processor or the dedicated BioHub.

\textbf{Heart Rate and Heart Rate Variability}
OmniBuds extract HR and HRV using the PPG signal, where HR is determined from peaks in the signal, and HRV is derived from inter-beat intervals. These computations occur on the BioHub, which also mitigates motion artefacts by combining PPG data with IMU readings. 

\textbf{Blood Oxygen Saturation}
Blood oxygen saturation (SpO$_{2}$) is calculated by analysing the ratio of the pulsatile and non-pulsatile components of red and infrared PPG signals. This computation is also handled by the BioHub, ensuring efficient processing while freeing the main MCU for other tasks.

\textbf{Respiration Rate}
OmniBuds estimate RR through PPG-based respiratory-induced intensity variation (RIIV). The signal is filtered to remove cardiac frequencies, and FFT is applied to determine the respiratory rate~\cite{romero2024optibreathe}. This process is performed by the main MCU using the PPG raw data through the Sensor Distribution module which manages the access to BioHub and PPG for both processed data (HR, HRV and SpO$_{2}$) and raw data.

\textbf{Blood Pressure}
OmniBuds support cuff-less blood pressure measurement using a multi-modal technique that captures the time difference between the S1 heart sound (detected by the in-ear microphone) and the PPG upstroke, known as vascular transit time (VTT)~\cite{truong2022non}. This, combined with ejection time (ET), enables systolic and diastolic blood pressure estimation through a personalised model. Future extensions could further enhance accuracy by leveraging OmniBuds' symmetrical hardware for multi-location sensing~\cite{balaji2023stereo}.

\subsubsection{Multi-modal Contextual Recognition}
In addition to monitoring all five vital signs and their derivatives, the sensor suite in OmniBuds enables the detection of typical motion-based contexts commonly tracked by earable devices. These include physical activity~\cite{kawsar2018esense, ma2021oesense}, head tracking~\cite{ferlini2019head}, head gestures~\cite{ma2021oesense}, and facial expressions~\cite{montanari2023earset, lee2019automatic}. While primarily driven by the integrated 9-axis IMU, more complex activities such as dietary monitoring~\cite{kawsar2018esense}, energy expenditure~\cite{gashi2022multidevice}, and mental fatigue assessment~\cite{kalanadhabhatta2021fatigueset} require combining data from multiple sensors, such as PPG and microphones.

What sets OmniBuds apart from previous platforms is its capability to run these diverse pipelines directly on-device, selecting the hardware component that is most suited to the task. For instance, low-power motion-based pipelines, such as activity recognition (e.g., walking, running) and head gestures detection (e.g., nodding, shaking), are efficiently processed directly on the IMU, minimising power consumption and reducing the load on the main MCU. Conversely, more complex tasks, such as dietary monitoring and fatigue estimation, can leverage the CNN accelerator to execute larger machine learning models, enabling the device to handle higher-capacity computations on-device with reduced latency~\cite{max7800_benchmark, moss2022ultra}.

\subsection{Future Applications}

\subsubsection{Emotion Recognition and Augmented Feedback}
OmniBuds could potentially integrate emotion recognition based on physiological changes, such as heart rate, body temperature, voice tone and facial expressions. For example, the device could monitor stress, anxiety and fatigue and adjust audio feedback accordingly, offering calming sounds or adjusting music tempo to match the user’s emotional state~\cite{butkow2024eartune}. This speculative application leverages multi-modal sensor data available on OmniBuds and their privacy-preserving design.

\subsubsection{Gesture-Based Interface for Cognitive Augmentation}
Using the IMU, OmniBuds could detect subtle head and facial movements, enabling gesture-based interfaces to control devices or interact with digital content hands-free. This could enhance AR/VR experiences or provide accessibility options for users with limited mobility.

\subsubsection{Personalised Fitness Coaching with Motion Analysis} OmniBuds' 9-axis IMU and CNN accelerator offer the potential for real-time motion analysis during workouts. By processing head movements and posture data directly on the device, OmniBuds could enable personalised fitness coaching, offering feedback on form and performance without needing external computation.

\subsubsection{Adaptive Personal Assistant Based on Physical and Cognitive State}
By monitoring a user’s physical and cognitive state, OmniBuds could adapt the behaviour of a personal assistant based on the user’s current conditions. For instance, when detecting fatigue or cognitive overload, the assistant could suggest breaks or adjust task complexity. This speculative application highlights the potential for a more responsive and adaptive interaction enabled by the OmniBuds' computational units.

\subsubsection{Acoustic Augmented Reality for Situational Awareness}
Using its dual microphones and Audio DSP, OmniBuds could enhance situational awareness by amplifying critical environmental sounds, such as approaching vehicles or alerts, while reducing background noise. This speculative application could augment safety in urban environments or outdoor activities. This shows the potential for real-time auditory augmentation offered by OmniBuds.

\section{Conclusion}
OmniBuds represent a significant advancement in the field of earable sensing and computing, offering a versatile platform that combines cutting-edge hardware with a flexible software architecture. With their integrated health monitoring, energy-efficient computing, and privacy-preserving design, OmniBuds open up new possibilities for researchers exploring diverse applications in wearable technology. As the platform continues to evolve, it has the potential to support groundbreaking research in areas such as health monitoring, cognitive interaction, and beyond. For updates and more information, visit \url{https://www.omnibuds.tech/}.

\bibliographystyle{ACM-Reference-Format}
\small
\balance
\bibliography{sample-base}

\end{document}